# Distributed Transmit Beamforming using Feedback Control

R. Mudumbai, *Student Member, IEEE*, J. Hespanha, *Senior Member, IEEE*, U. Madhow, *Fellow, IEEE* and G. Barriac, *Member, IEEE*

***Abstract*— A simple feedback control algorithm is presented for distributed beamforming in a wireless network. A network of wireless sensors that seek to cooperatively transmit a common message signal to a Base Station (BS) is considered. In this case, it is well-known that substantial energy efficiencies are possible by using distributed beamforming. The feedback algorithm is shown to achieve the carrier phase coherence required for beamforming in a scalable and distributed manner. In the proposed algorithm, each sensor independently makes a random adjustment to its carrier phase. Assuming that the BS is able to broadcast one bit of feedback each timeslot about the change in received signal to noise ratio (SNR), the sensors are able to keep the favorable phase adjustments and discard the unfavorable ones, asymptotically achieving perfect phase coherence. A novel analytical model is derived that accurately predicts the convergence rate. The analytical model is used to optimize the algorithm for fast convergence and to establish the scalability of the algorithm.**

This work was supported by the National Science Foundation under grants CCF-0431205, ANI-0220118 and EIA-0080134, by the Office of Naval Research under grant N00014-03-1-0090, and by the Institute for Collaborative Biotechnologies through grant DAAD19-03-D-0004 from the U.S. Army Research Office.

R. Mudumbai, J. Hespanha and U. Madhow are with the Department of Electrical and Computer Engineering, University of California Santa Barbara

G. Barriac is with Qualcomm Inc., San Diego, CA

## I. Introduction

Energy efficient communication is important for wireless ad-hoc and sensor networks. We consider the problem of cooperative communication in a sensor network, where there are multiple transmitters (e.g., sensor nodes) seeking to transmit a common message signal to a distant Base Station receiver (BS). In particular, we investigate *distributed beamforming*, where multiple transmitters coordinate their transmissions to combine coherently at the intended receiver. With beamforming, the signals transmitted from each antenna undergo constructive interference at the receiver, the multiple transmitters acting as a *virtual antenna array*. Thus, the received signal magnitude increases in proportion to number of transmitters $N$, and the SNR increases proportional to $N^2$, whereas the total transmit power only increases proportional to $N$. This $N$-fold increase in energy efficiency, however, requires precise control of the carrier phases at each transmitter in order that the transmitted signals arrive in phase at the receiver. In this paper, we propose a feedback control protocol for achieving such phase coherence. The protocol is based on a fully distributed iterative algorithm, in which each transmitter independently adjusts its phase by a small amount in a manner depending on a single bit of feedback from the BS. The algorithm is scalable, in that convergence to phase coherence occurs in a time that is linear in the number of cooperating transmitters.

Prior work on cooperative communication mainly focuses on exploiting spatial diversity for several wireless relaying and networking problems [1], [2]. Such distributed diversity methods require different transmitters to transmit information on orthogonal channels, which are then combined at the receiver. The resulting diversity gains could be substantial in terms of smoothing out statistical fluctuations in received power due to fading and shadowing environments. However, unlike distributed beamforming, distributed diversity does not provide a gain in energy efficiency in terms of *average* received power, which simply scales with the transmitted power. On the other hand, the coherent combining of signals at the receiver due to distributed beamforming also provides diversity gains.

Recent papers discussing potential gains from distributed beamforming include [3], which investigates the use of beamforming for relay under ideal coherence at the receiver, and [4], which shows that even partial phase synchronization leads to significant increase in power efficiency in wireless ad hoc networks. The beam patterns resulting from distributed beamforming using randomly placed nodes are investigated in [5]. However, the technical feasibility of distributed beamforming is not investigated in the preceding papers. In our prior work [6], [7], we recognized that the key technical bottleneck in distributed beamforming is carrier phase synchronization across cooperating nodes. We presented a protocol in which the nodes first establish a common carrier phase reference using a *master-slave* architecture, thus providing a direct emulation of a centralized multi-antenna system. This is a challenging problem, because even small timing errors lead to large phase errors at the carrier frequencies of interest. Once phase synchronization is achieved, reciprocity was proposed as a means of measuring the channel phase response to the BS. In this paper, we present an alternative method of achieving coherent transmission iteratively using a simple feedback control algorithm, which removes the need for explicit estimation of the channel to the BS, and greatly reduces the level of coordination required among the sensors.

Other related work on synchronization in sensor networks is based on pulse-coupled oscillator networks [8] and biologically inspired (firefly synchronization) [9] methods. These methods are elegant, robust and suitable for distributed implementation, however they are limited by assumptions of zero propagation delay and the requirement of mesh-connectivity, and are not suitable for carrier phase synchronization.

We consider the following model to illustrate our ideas. The protocol is initialized by each sensor transmitting a common message signal modulated by a carrier with an arbitary phase offset. (This phase offset is a result of unknown timing synchronization errors, and is therefore unknown.) When the sensors' wireless channel is linear and time-invariant, the received signal is the message signal modulated by an effective carrier signal that is the phasor sum of the channel-attenuated carrier signals of the individual sensors. At periodic intervals, the BS broadcasts a single bit of feedback to the sensors indicating whether or not the received SNR level increased in the preceding interval. Each sensor introduces an independent random perturbation of their transmitted phase offset. When this results in increased total SNR compared to the previous time intervals (as indicated by feedback from the BS), the new phase offset is set equal to the perturbed phase by each sensor; otherwise, the new phase offset is set equal to the phase prior to the perturbation. Each sensor then introduces a new random perturbation, and the process continues. We show that this procedure asymptotically converges to perfect coherence of the received signals, and provide a novel analysis that accurately predicts the rate of convergence. We verify the analytical model using Monte-Carlo simulations, and use it to optimize the convergence rate of the algorithm.

The rest of this paper is organized as follows. Section II describes our communication model for the sensor network. A feedback control protocol for distributed beamforming is described in Section III-A and its asymptotic convergence is shown in Section III-B. Section IV describes an analytical model to characterize the convergence behavior of the protocol. Some analytical and simulation results are presented in Section V. Section V-A presents an optimized version of the feedback control protocol. Sections V-B and V-C present some results on scalability, and the effect of time-varying channels respectively. Section VI concludes the paper with a short discussion of open issues.

## II. Communication Model for a Sensor Network

We consider a system of $N$ sensors transmitting a common message signal $m(t)$ to a receiver. Each sensor is power constrained to a maximum transmit power of $P$. The message $m(t)$ could represent raw measurement data, or it could be a waveform encoded with digital data. We now list the assumptions in this model.

1) The sensors communicate with the receiver over a narrowband wireless channel at some carrier frequency, $f_c$. In particular, the message bandwidth $B < W_c$, where $B$ is the bandwidth of $m(t)$ and $W_c$ is the coherence bandwidth of each sensor's channel. This means that each sensor has a flat-fading channel to the receiver. Therefore the sensor $i$'s channel can be represented by a complex scalar gain $h_i$.
2) Each sensor has a local oscillator synchronized to the carrier frequency $f_c$ i.e. carrier drift is small. One way to ensure this is to use Phase-Locked Loops (PLLs) to synchronize to a reference tone transmitted by a designated master sensor as in [6]. In this paper, we use complex-baseband notation for all the transmitted signals referred to the common carrier frequency $f_c$.
3) The local carrier of each sensor $i$ has an unknown phase offset, $\gamma_i$ relative to the receiver's phase reference. Note that synchronization using PLLs still results in independent random phase offsets $\gamma_i = (2\pi f_c \tau_i \mod 2\pi)$, because of timing synchronization errors $\tau_i$ that are fundamentally limited by propagation delay effects.
4) The sensors' communication channel is time-slotted with slot length $T$. The sensors only transmit at the beginning of a slot. This requires some coarse timing synchronization: $\tau_i \ll T$ where $\tau_i$ is the timing error of sensor i.
5) Timing errors among sensors are small compared to a symbol interval (a "symbol interval" $T_s$ is nominally defined as inverse bandwidth: $T_s = \frac{1}{B}$). For a digitally modulated message signal $m(t)$, this means that Inter Symbol Interference (ISI) can be neglected.
6) The channels $h_i$ are assumed to exhibit slow-fading, i.e. the channel gains stay roughly constant for several time-slots. In other words $T_s \ll T \ll T_c$, where $T_c$ is the coherence time of the sensor channels.

**Distributed transmission model:** The communication process begins with the receiver broadcasting a signal to the sensors to transmit their measured data. The sensors then transmit the message signal at the next time-slot. Specifically, each sensor transmits: $s_i(t) = A\cdot g_i m(t-\tau_i)$, where $\tau_i$ is the timing error of sensor $i$, $A \propto \sqrt{P}$ is the amplitude of the transmission, and $g_i$ is a complex amplification performed by sensor $i$. Our objective is to choose $g_i$ to achieve optimum received SNR, given transmit power constraint of $P$ on each sensor. For simplicity, we write $h_i = a_i e^{j\psi_i}$ and $g_i = b_i e^{j\theta_i}$, where $b_i \leq 1$ to satisfy the power constraint. Then the received signal is:

$$r(t) = \sum_{i=1}^{N} h_i s_i(t) e^{j\gamma_i} + n(t) \tag{1}$$

$$= A \sum_{i=1}^{N} h_i g_i e^{j\gamma_i} m(t-\tau_i) + n(t)$$

$$= A \sum_{i=1}^{N} a_i b_i e^{j(\gamma_i+\theta_i+\psi_i)} m(t-\tau_i) + n(t). \tag{2}$$

In the frequency domain, this becomes:

$$R(f) = A \sum_{i=1}^{N} a_i b_i e^{j(\gamma_i+\theta_i+\psi_i)} M(f) e^{-jf\tau_i} + N(f)$$

$$\approx A\cdot M(f) \sum_{i=1}^{N} a_i b_i e^{j(\gamma_i+\theta_i+\psi_i)} + N(f), \tag{3}$$

where $n(t)$ is the additive noise at the receiver and $N(f)$ is its Fourier transform over the frequency range $\left|f\right| \leq \frac{B}{2}$.

In (1), the phase term $\gamma_i$ accounts for the phase offset in sensor $i$. In (3), we set $e^{-jf\tau_i} \approx 1$ because $f\tau_i \leq B\tau_i \equiv \frac{\tau_i}{T_s} \ll 1$. Equation (3) motivates a figure of merit for the beamforming gain:

$$G = \left|\sum_{i=1}^{N} a_i b_i e^{j(\gamma_i+\theta_i+\psi_i)}\right| \tag{4}$$

which is proportional to the square-root of received SNR.

Note that $b_i \leq 1$, in order to satisfy the power constraint on sensor $i$. From the Cauchy-Schwartz Inequality, we can see that to maximize $G$, it is necessary that the received carrier phases $\Phi_i \doteq \gamma_i + \theta_i + \psi_i$, are all equal:

$$G = \left|\sum_{i=1}^{N} a_i b_i e^{j\Phi_i}\right|$$

$$\leq \left|\sum_{i=1}^{N} a_i e^{j\Phi_i}\right|$$

$$= \left|\sum_{i=1}^{N} \left(\sqrt{a_i}\right)\left(\sqrt{a_i} e^{j\Phi_i}\right)\right| \tag{5}$$

$$\leq G^{opt} \equiv \left(\sum_{i=1}^{N} a_i\right), \text{ with equality if and only if } \Phi_i = \Phi_j \text{ and } b_i = 1 \tag{6}$$

However sensor $i$ is unable to estimate either $\gamma_i$ or $\psi_i$ because of the lack of a common carrier phase reference. In the rest of this paper, we propose and analyze a feedback control technique for sensor $i$ to dynamically compute the optimal value of $\theta_i$ so as to achieve the condition for equality in (6).

## III. Feedback Control Protocol

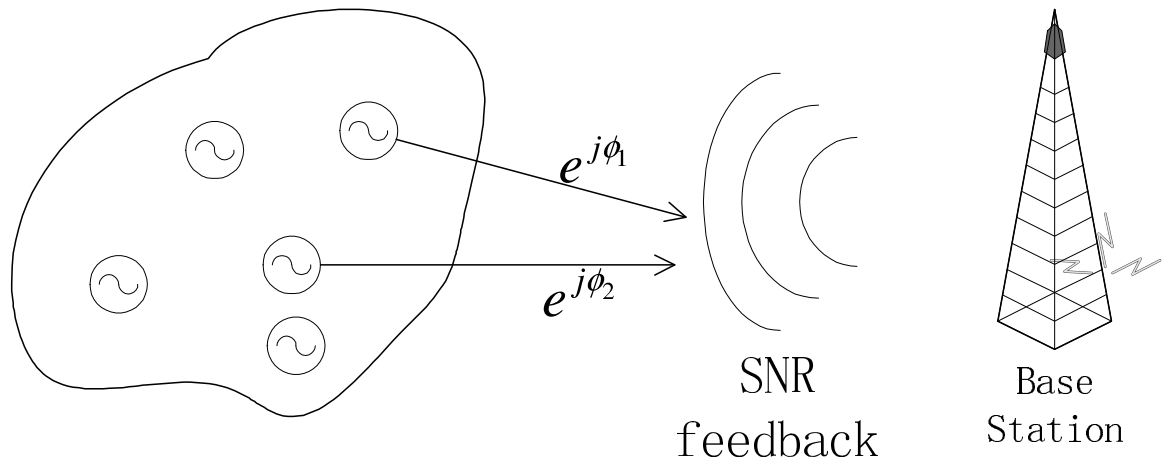

Fig. 1. Phase synchronization using receiver feedback

Fig. 1 illustrates the process of phase synchronization using feedback control. In this section, we describe the feedback control algorithm, and prove its asymptotic convergence.

### A. Description of Algorithm

The protocol for distributed beamforming works as follows: each sensor starts with an arbitrary (unsynchronized) phase offset $\gamma_i$. In each time-slot, the sensor applies a random perturbation to $\theta_i$ and observes the resulting received signal strength $y[n]$ through feedback. The objective is to adjust its phase to maximize $y[n]$ through coherent combining at the receiver. Each phase perturbation is a guess by each sensor about the correct phase adjustment required to increase the overall received signal strength. If the received SNR is found to increase as a result of this perturbation, the sensor adds the appropriate phase offset, and repeats the process. This works like a distributed, randomized gradient search procedure, and eventually converges to the correct phase offsets for each sensor to achieve distributed beamforming. Fig. 2 shows the convergence to beamforming with $N = 10$ sensors.

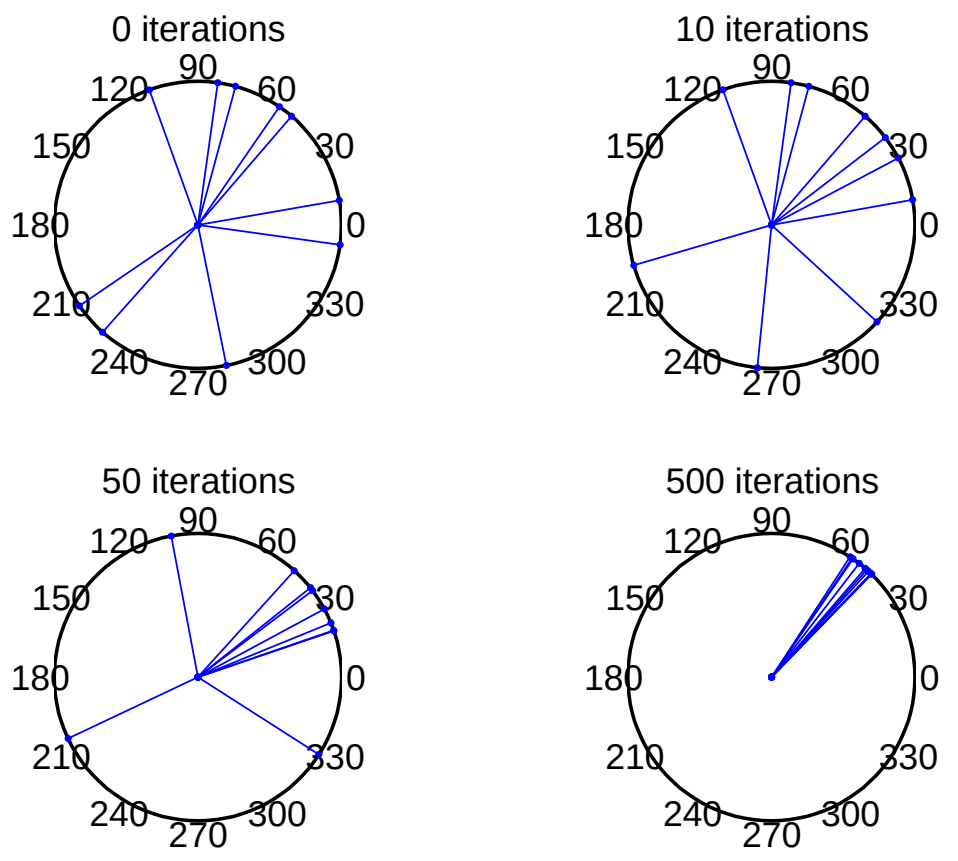

Fig. 2. Convergence of feedback control algorithm

Let $n$ denote the time-slot index and $\mathcal{Y}[n]$ the amplitude of the received signal in time-slot $n$. From (3), we have: $\mathcal{Y}[n] \propto \left|\sum_i a_i e^{j\Phi_i[n]}\right|$ where $\Phi_i[n]$ is the received signal phase corresponding to sensor $i$. We set the proportionality constant to unity for simplicity of analysis. At each time instant $n$, let $\theta_i[n]$ be the best known carrier phase at sensor $i$ for maximum received SNR. Each sensor uses the distributed feedback algorithm to dynamically adjust $\theta_i[n]$ to satisfy (6) asymptotically. The algorithm works as follows.

Initially the phases are set to zero: $\theta_i[1] = 0$. At each time-slot $n$, each sensor $i$ applies a random phase perturbation $\delta_i[n]$ to $\theta_i[n]$ for its transmission. As a result, the received phase is given by: $\Phi_i[n] = \gamma_i + \theta_i[n] + \delta_i[n] + \psi_i$. The BS measures $\mathcal{Y}[n]$ and keeps a record of the highest observed signal strength $\mathcal{Y}_{best}[n] = \max_{k<n} \mathcal{Y}[k]$ in all previous timeslots. At the end of each timeslot, the BS broadcasts a one-bit feedback message that indicates whether the received signal strength of the preceding timeslot was higher than the previous highest signal strength. Depending on the feedback message, each sensor $i$ updates its phase according to:

$$\theta_i[n+1] = \begin{cases} \theta_i[n] + \delta_i[n] & \mathcal{Y}[n] > \mathcal{Y}_{best}[n] \\ \theta_i[n] & \text{otherwise.} \end{cases} \tag{7}$$

Simultaneously, The BS also updates its highest received signal strength:

$$\mathcal{Y}_{best}[n+1] = \max\Big(\mathcal{Y}_{best}[n], \mathcal{Y}[n]\Big) \tag{8}$$

This has the effect of retaining the phase perturbations that increase SNR and discarding the unfavorable ones that do not increase SNR. The sensors and the BS repeat the same procedure in the next timeslot.

The random perturbation $\delta_i[n]$ is chosen independently across sensors from a probability distribution $\delta_i[n] \sim f_\delta(\delta_i)$, where the density function $f_\delta(\delta_i)$ is a parameter of the protocol. In this paper, we consider primarily two simple distributions for $f_\delta(\delta_i)$: (i) the two valued distribution where $\delta_i = \pm\delta_0$ with probability $0.5$, and (ii) the uniform distribution where $\delta_i \sim \text{uniform}[-\delta_0, \delta_0]$. We allow for the possibility that the distribution $f_\delta(\delta_i)$ dynamically changes in time.

It follows from (7) that if the algorithm were to be terminated at timeslot $n$, the best achievable signal strength using the feedback information received so far, is equal to $\mathcal{Y}_{best}[n]$, which correspond to sensor $i$ transmitting with the phase $\theta_i[n]$.

$$\mathcal{Y}_{best}[n] \equiv |\sum_i a_i e^{\Phi_i[n]}|$$
$$\text{where } \Phi_i[n] = \gamma_i + \theta_i[n] + \psi_i \qquad (9)$$

## B. Asymptotic Coherence

We now show that the feedback control protocol outlined in Section III-A asymptotically achieves phase coherence for any initial values of the phases $\Phi_i$. Let $\bar{\Phi}$ denote the vector of the received phase angles $\Phi_i$. We define the function $\mathrm{Mag}(\bar{\Phi})$ to be the received signal strength corresponding to received phase $\bar{\Phi}$:

$$\mathrm{Mag}(\bar{\Phi}) \doteq |\sum_i a_i e^{j\Phi_i}| \qquad (10)$$

Phase coherence means $\Phi_i = \Phi_j = \Phi_{const}$, where $\Phi_{const}$ is an arbitrary phase constant. In order to remove this ambiguity, it is convenient to work with the rotated phase values $\phi_i = \Phi_i - \Phi_0$, where $\Phi_0$ is a constant chosen such that the phase of the total received signal is zero. This is just a convenient shift of the receiver's phase reference and as (10) shows, such a shift has no impact on the received signal strength:

$$\mathrm{Mag}(\bar{\phi}) \equiv \mathrm{Mag}(\bar{\Phi}) \qquad (11)$$

We interpret the feedback control algorithm as a discrete-time random process $\mathcal{Y}_{best}[n]$ in the state-space of $\bar{\phi}$, the state-space being the $N$-dimensional space of the phases $\phi_i$ constrained by the condition that the phase of the received signal is zero. We observe that the sequence $\{\mathcal{Y}_{best}[n]\}$ is monotonically non-decreasing, and is upperbounded by $G^{opt}$ as shown in (5). Therefore each realization of $\{\mathcal{Y}_{best}[n]\}$ is always guaranteed to converge to some limit $G_0 \leq G^{opt}$. Furthermore $G_0 = \lim_{n\to\infty} \mathcal{Y}_{best}[n] \equiv \sup_{n\to\infty} \mathcal{Y}_{best}[n]$ i.e. the limit $G_0$ of the sequence $\{\mathcal{Y}_{best}[n]\}$ is the same as its Least Upper Bound [10]. In general the limit $G_0$ would depend on the starting phase angles $\bar{\phi}$. We now provide an argument that shows (under mild conditions on the probability density function $f_\delta(\delta_i)$), that in fact $\{\mathcal{Y}_{best}[n]\}$ converges to the constant $G^{opt}$ *with probability 1* for arbitrary starting phases $\bar{\phi}$. The following proposition will be needed to establish the convergence.

**Proposition 1:** Consider a distribution $f_\delta(\delta_i)$ that has non-zero support in an interval $(-\delta_0, \delta_0)$. Given any $\bar{\phi} \neq \bar{0}$, and $\mathrm{Mag}(\bar{\phi}) < G^{opt} - \epsilon$, where $\epsilon > 0$ is arbitrary, there exist constants $\epsilon_1 > 0$ and $\rho > 0$ such that $Prob(\mathrm{Mag}(\bar{\phi} + \bar{\delta}) - \mathrm{Mag}(\bar{\phi}) > \epsilon_1) > \rho$.

*Proof.* For the class of distributions $f_\delta(\delta_i)$ that we consider, the probability of choosing $\delta_i$ in any finite interval $I \subset (-\delta_0, \delta_0)$ is non-zero. One example of such a class of distributions is $f_\delta(\delta_i) \sim \mathrm{uniform}[-\delta_0, \delta_0]$.

Recall that the phase reference is chosen such that the total received signal $\sum_i a_i e^{j\phi_i}$ has zero phase. First we sort all the phases $\phi_i$ in the vector $\bar{\phi}$ in the descending order of $|\phi_i|$ to get the sorted phases $\phi_i^*$ satisfying $|\phi_1^*| > |\phi_2^*| > ... > |\phi_N^*|$, and the corresponding sorted channel gains $a_i^*$. We use the condition $\mathrm{Mag}(\bar{\phi}) < G^{opt} - \epsilon$ to get:

$$\cos(\phi_1^*) \sum_i a_i^* < \sum_i a_i^* \cos(\phi_i^*) \leq G^{opt} - \epsilon$$
$$\phi_1^* > \phi_\epsilon \doteq \cos^{-1}\Big(\frac{G^{opt} - \epsilon}{\sum_i a_i^*}\Big) \qquad (12)$$

Now we choose a phase perturbation $\delta_1$ that decreases $|\phi_1^*|$. This makes the most mis-aligned phase in $\bar{\phi}$ closer to the received signal phase, and thus increases the magnitude of the received signal. If $\phi_1^* > 0$, then we need to choose a $\delta_1 < 0$, whereas if $\phi_1^* < 0$, we need $\delta_1 > 0$. In the following, we assume that $\phi_1^* > 0$ and $\phi_\epsilon > \delta_0$. The argument below does not depend on these assumptions, and can be easily modified for the other cases. Consider $\delta_1 \in (-\delta_0, -\frac{\delta_0}{2})$. This is an interval in which $f_\delta(\delta_1)$ is non-zero , therefore there is a non-zero probability $\rho_1 > 0$ of choosing such a $\delta_1$. We have:

$$a_1^* \cos(\phi_1^* + \delta_1) - a_1^* \cos(\phi_1^*) > 2\epsilon_1$$
$$\text{where } \epsilon_1 \doteq \frac{a_1^* \sin(\phi_\epsilon^* - \frac{\delta_0}{2})\delta_0}{4} \quad (13)$$

We observe that $\epsilon_1$ and $\rho_1$ do not dependent on $\bar{\phi}$.

The perturbation $\delta_1$ by itself will achieve a non-zero increase in total received signal, provided that the other phases $\phi_i^*$ do not get too mis-aligned by their respective $\delta_i$:

$$\begin{aligned} \mathrm{Mag}(\bar{\phi} + \bar{\delta}) - \mathrm{Mag}(\bar{\phi}) &= \sum_i a_i^* \big(\cos(\phi_i^* + \delta_i) - \cos(\phi_i^*)\big) \\ &= a_1^* \big(\cos(\phi_1^* + \delta_1) - \cos(\phi_1^*)\big) + \sum_{i>1} a_i^* \big(\cos(\phi_i^* + \delta_i) - \cos(\phi_i^*)\big) \\ &> 2\epsilon_1 + \sum_{i>1} a_i^* \big(\cos(\phi_i^* + \delta_i) - \cos(\phi_i^*)\big) \end{aligned} \quad (14)$$

We note that since $\mathrm{Mag}(\bar{\phi})$ is continuous in each of the phases $\phi_i^*$, we can always find a $\epsilon_i > 0$ to satisfy:

$$\Big| a_i^* \big(\cos(\phi_i^* + \delta_i) - \cos(\phi_i^*)\big) \Big| < \frac{\epsilon_1}{N-1}, \forall |\delta_i| < \epsilon_i \quad (15)$$

In particular the choice $\epsilon_i \doteq \frac{\epsilon_1}{a_i^*(N-1)}$, satisfies (15), and this choice of $\epsilon_i$ is independent of $\bar{\phi}$. With the $\delta_i$'s chosen to satisfy (15), we have:

$$-\epsilon_1 < \sum_{i>1} a_i^* \big(\cos(\phi_i^* + \delta_i) - \cos(\phi_i^*)\big) < \epsilon_1 \quad (16)$$

Since $f_\delta(\delta_i)$ has non-zero support in each of the non-zero intervals $(-\epsilon_i, \epsilon_i)$, the probability $\rho_i$ of choosing $\delta_i$ to satisfy (15) is non-zero, i.e. $\rho_i > 0$, which is independent of $\bar{\phi}$. Finally, we recall that each of the $\delta_i$ are chosen independently, and therefore with probability $\rho = \prod_i \rho_i > 0$, it is possible to find $\delta_1$ to satisfy (13) and $\delta_i, i > 1$ to satisfy (15). For $\bar{\delta}$ chosen as above, $\mathrm{Mag}(\bar{\phi} + \bar{\delta}) - \mathrm{Mag}(\bar{\phi}) > \epsilon_1$, and therefore Proposition 1 follows. □

**Theorem 1:** For the class of distributions $f_\delta(\delta_i)$ considered in Proposition 1, starting from an arbitrary $\bar{\phi}$, the feedback algorithm converges to perfect coherence of the received signals almost surely, i.e. $\mathcal{Y}_{best}[n] \to G^{opt}$ or equivalently $\bar{\phi}[n] \to \bar{0}$ (i.e. $\phi_i[n] \to 0, \forall i$) *with probability 1.*

*Proof:* We wish to show that the sequence $\mathcal{Y}_{best}[n] = \mathrm{Mag}(\bar{\phi}[n]) \to G^{opt}$ given an arbitrary $\bar{\phi}[1] = \bar{\phi}$. Consider an arbitrarily small $\epsilon > 0$ and define $T_\epsilon(\bar{\phi})$ as the first timeslot when the received signal exceeds $G^{opt} - \epsilon$.

By definition if $n < T_\epsilon(\bar{\phi})$, then $\mathcal{Y}_{best}[n] = \mathrm{Mag}(\bar{\phi}[n]) < G^{opt} - \epsilon$, and by Proposition 1, $Prob(\mathcal{Y}_{best}[n+1] - \mathcal{Y}_{best}[n] > \epsilon_1) > \rho$ for some constants $\epsilon_1 > 0$ and $\rho > 0$. We have:

$$E\Big(\mathcal{Y}_{best}[n+1] - \mathcal{Y}_{best}[n]\Big) > \epsilon_1 \rho, \forall n < T_\epsilon(\bar{\phi}) \quad (17)$$

Using (17) we have:

$$\begin{aligned} G^{opt} &\geq \mathcal{Y}_{best}[n+1] \\ &= \mathcal{Y}_{best}[1] + \sum_{k=1}^{n} \Big(\mathcal{Y}_{best}[k+1] - \mathcal{Y}_{best}[k]\Big) \\ &> \sum_{k=1}^{n} \Big(\mathcal{Y}_{best}[k+1] - \mathcal{Y}_{best}[k]\Big) \end{aligned} \quad (18)$$

Taking expectation we have:

$$\begin{aligned} G^{opt} &> E\Big(\sum_{k=1}^{n} (\mathcal{Y}_{best}[k+1] - \mathcal{Y}_{best}[k])\Big) \\ &> Prob\Big(T_\epsilon(\bar{\phi}) > n\Big) E\Big(\sum_{k=1}^{n} (\mathcal{Y}_{best}[k+1] - \mathcal{Y}_{best}[k]) \Big| T_\epsilon(\bar{\phi}) > n\Big) \\ &> Prob\Big(T_\epsilon(\bar{\phi}) > n\Big) n \epsilon_1 \rho \end{aligned} \quad (19)$$

where we obtained (19) by using (17). Therefore we have $Prob(T_\epsilon(\bar{\phi}) > n) < \frac{1}{n} \frac{G^{opt}}{\epsilon_1 \rho} \to 0$, as $n \to \infty$. Since this is true for an arbitrarily small $\epsilon$, we have shown that $\mathcal{Y}_{best}[n] \to G^{opt}$ and $\bar{\phi}[n] \to \bar{0}$ almost surely. □

## IV. Analytical Model for Convergence

The analysis in Section III-B shows that the feedback control algorithm of Section III-A asymptotically converges for a large class of distributions $f_\delta(\delta_i)$; however it provides no insight into the rates of convergence. We now derive an analytical model based on simple, intuitive ideas that predicts the convergence behavior of the protocol accurately. We then use this analytical model, to optimize $f_\delta(\delta_i)$ for fast convergence.

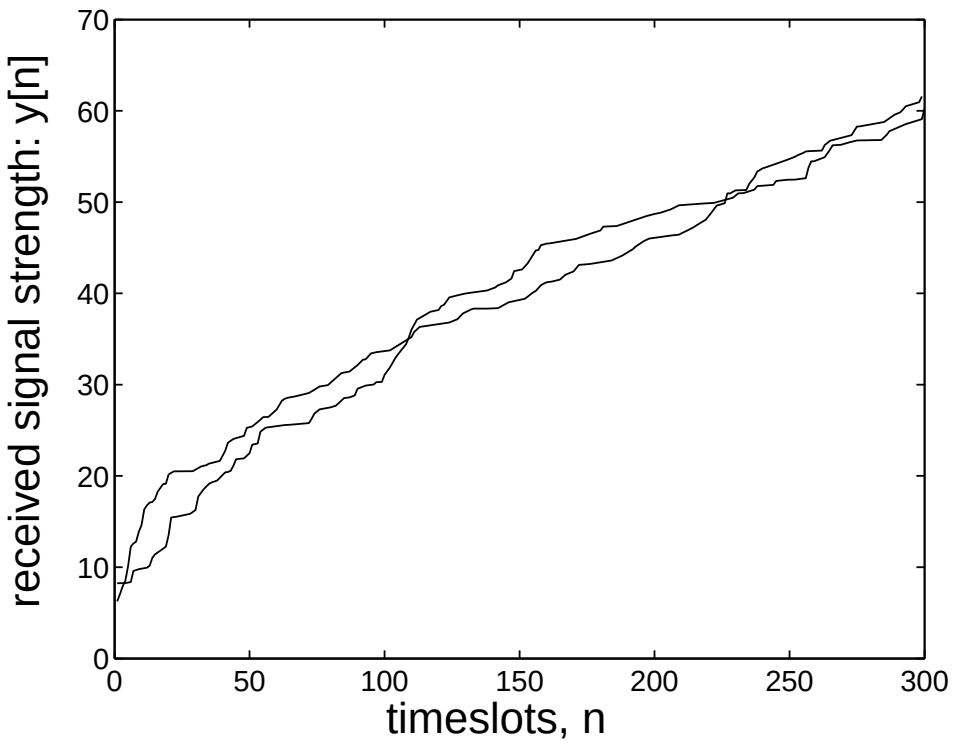

Fig. 3. Motivating the Analytical Model: two simulated instances with $N = 100, f_\delta(\delta_i) \sim \text{uniform}[-\frac{\pi}{20}, \frac{\pi}{20}]$.

### A. Derivation of Analytical Model

The basic idea behind our analytical model is that the convergence rate of typical realizations of $\mathcal{Y}_{best}[n]$ is well-modeled by computing the *expected increase* in signal strength at each time-interval given a distribution $f_\delta(\delta_i)$. This is illustrated in Fig. 3, where we show two separate realizations of $\mathcal{Y}_{best}[n]$ from a Monte-Carlo simulation of the feedback algorithm.

We define the averaged sequence $y[n]$ recursively as the conditional value of $\mathcal{Y}_{best}[n+1]$ given $\mathcal{Y}_{best}[n]$:

$$y[1] = E\Big(\text{Mag}(\bar{\phi}[1])\Big) \quad (20)$$

$$y[n+1] = E_{\bar{\delta}[n]}\Big(\mathcal{Y}_{best}[n+1]\Big|\mathcal{Y}_{best}[n] = y[n]\Big) \quad (21)$$

The initial value $y[1]$ in (20) is set under the assumption that the received phases $\bar{\phi}[1]$ are randomly distributed in $[0, 2\pi)$. For subsequent timeslots, $\mathcal{Y}_{best}[n+1]$ in (21) is conditioned on $\mathcal{Y}_{best}[n]$ but the phase vector $\bar{\phi}[n]$ is not known. Some remarks are in order regarding this definition, particularly the relationship of $y[n]$ with the (unconditionally) averaged $\mathcal{Y}_{best}[n]$. Let

$$y[n+1] = F\big(y[n]\big), \text{ where } F(y) \doteq E\big(\mathcal{Y}_{best}[n+1]\big|\mathcal{Y}_{best}[n] = y\big) \quad (22)$$

Consider:

$$\begin{aligned} E\big(\mathcal{Y}_{best}[n+1]\big) &= E_{\mathcal{Y}_{best}[n]}\Big(E\big(\mathcal{Y}_{best}[n+1]\big|\mathcal{Y}_{best}[n]\big)\Big) \\ &= E_{\mathcal{Y}_{best}[n]}\Big(F\big(\mathcal{Y}_{best}[n]\big)\Big) \\ &\approx F\Big(E\big(\mathcal{Y}_{best}[n]\big)\Big) \equiv y[n+1] \quad (23) \end{aligned}$$

In most cases, the function $F(y)$ is concave, and therefore (by Jensen's Inequality) the approximation in (23) represents an overestimate of the unconditional average of $\mathcal{Y}_{best}[n+1]$. Also in different instances of the algorithm, we would expect to see different random evolutions of $\bar{\phi}[n]$ and $\mathcal{Y}_{best}[n]$ with time, and an averaged quantity only provides partial information about the convergence rate. Fortunately, as Fig. 3 shows, even over multiple instances of the algorithm, the convergence rate remains highly predictable, and the average characterizes the actual convergence reasonably well. Since the variation of the random $\mathcal{Y}_{best}[n]$ around its average value is small, the approximation in (23) also works well. Our goal is to compute $F(y)$ as defined in (22).

Note that while (22) is conditioned on $\mathcal{Y}_{best}[n]$ being known, the phase vector $\bar{\phi}[n]$ is unknown. As $\mathcal{Y}_{best}[n]$ increases, the phases $\phi_i[n]$ become increasingly clustered together, however their exact values are determined by their initial values, and the random perturbations from previous time-slots. In order to compute the expectation in (22), we need some information about $\bar{\phi}[n]$.

We show in Section IV-B that the phases $\phi_i[n]$ can be accurately modeled as clustered together according to a statistical distribution that is determined parametrically as a function of $\mathcal{Y}_{best}[n]$ alone. This is analogous to the technique in equilibrium statistical mechanics, where the individual positions and velocities of particles in an ensemble is unknown, but accurate macroscopic results are obtained by modeling the kinetic energies as following the Boltzmann distribution, which is fully determined by a single parameter (the average kinetic energy or the temperature). In our case, $\phi_i[n]$ are modeled as independent and identically distributed (for all $i$) according to a distribution satisfying the constraint:

$$y[n] = \sum_i a_i \cos\phi_i \equiv NE\big(a_i\big)E_{\phi_i}\big(\cos\phi_i\big) \qquad (24)$$

Therefore, even though the individual $\phi_i$ are unknown, we can compute all aggregate functions of $\bar{\phi}$ using this distribution, as if the $\phi_i$ are known. This is an extremely powerful tool, and we now use it to compute $F(y)$ treating $\bar{\phi}[n]$ as a given. Section IV-B completes the computation by deriving the distribution used to specify $\bar{\phi}[n]$ given $\mathcal{Y}_{best}[n]$.

From the condition $\mathcal{Y}_{best}[n] = y[n]$, we have:

$$y[n] = \sum_i a_i e^{j\phi_i} = \sum_i a_i \cos\phi_i \qquad (25)$$

where we used the fact that the imaginary part of the received signal is zero for our choice of phase reference. We have the following expressions (omitting the time-index on $\bar{\phi}[n]$ and $\bar{\delta}[n]$ for convenience):

$$\mathcal{Y}_{best}[n+1] = \begin{cases} \mathrm{Mag}(\bar{\phi}+\bar{\delta}) & \text{if } \mathrm{Mag}(\bar{\phi}+\bar{\delta}) > y[n] \\ y[n] & \text{otherwise.} \end{cases} \qquad (26)$$

We now express $\mathrm{Mag}(\bar{\phi}+\bar{\delta})$ as a sum of i.i.d. terms from each sensor, and invoke the Central Limit Theorem (CLT).

$$\mathrm{Mag}(\bar{\phi}+\bar{\delta}) = \Big|\sum_i a_i e^{j\phi_i + j\delta_i}\Big| \qquad (27)$$

$$= \Big|\sum_i a_i\big(\cos\phi_i\cos\delta_i - \sin\phi_i\sin\delta_i\big) + j\sum_i a_i\big(\cos\phi_i \sin\delta$$

$$= \Big|\big(C_\delta y[n] + x_1\big) + jx_2\Big|, \qquad (28)$$

$$\text{where } C_\delta = E_\delta\big(\cos\delta_i\big), \qquad (29)$$

$$x_1 = \sum_i a_i\Big(\cos\phi_i\big(\cos\delta_i - C_\delta\big) - \sin\phi_i\sin\delta_i\Big), \qquad (30)$$

$$x_2 = \sum_i a_i\Big(\cos\phi_i\sin\delta_i + \sin\phi_i\cos\delta_i\Big) \qquad (31)$$

The random variables $x_1, x_2$ are illustrated in Fig. 4.

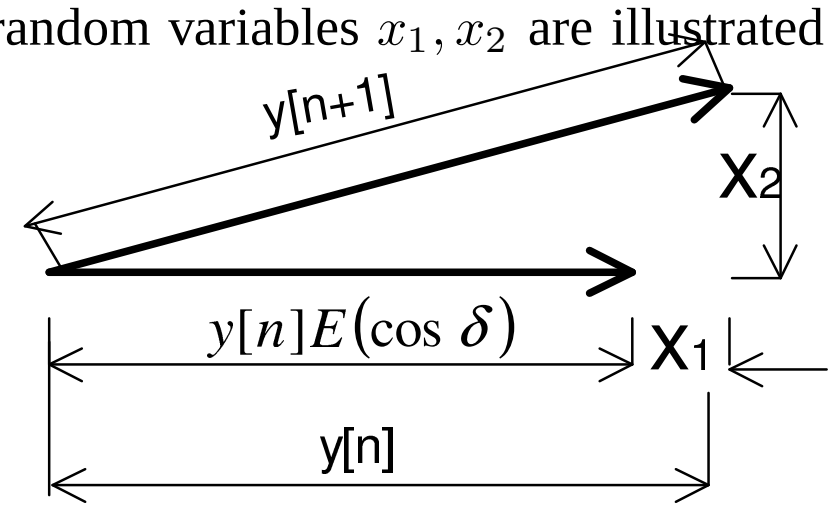

Fig. 4. Perturbation in the total received signal.

Both $x_1$ and $x_2$ as defined in (28) are linear combinations of iid random variables, $\sin\delta_i$ and $\cos\delta_i$. Therefore as the number of sensors $N$ increases, these random variables can be well-modeled as Gaussian, as per the CLT [11]. Futhermore, $x_1, x_2$ are zero-mean random variables, and their respective variances $\sigma_1^2, \sigma_2^2$ are related by:

$$\sigma_1^2 = \frac{1}{2}\sum_i a_i^2\Big((1-C_\delta^2) - \cos(2\phi_i)(C_\delta^2 - C_{2\delta})\Big)$$

$$\sigma_2^2 = \frac{1}{2}\sum_i a_i^2\Big(1 - \cos(2\phi_i)C_{2\delta}\Big)$$

$$\text{where } C_{2\delta} = E_\delta\big(\cos(2\delta_i)\big) \qquad (32)$$

With these simplifications, the statistics of $y[n+1]$ only depends on the density function $f_\delta(\delta_i)$ through $C_\delta$ and $C_{2\delta}$. We have the following proposition.

**Proposition 2:** Assuming that the CLT applies for random variable $x_1$, the expected value of the received signal strength is given by:

$$y[n+1] \approx y[n]\Big(1 - p\cdot(1-C_\delta)\Big) + \frac{\sigma_1}{\sqrt{2\pi}} e^{-\frac{(y[n](1-C_\delta))^2}{2\sigma_1^2}} \tag{33}$$

$$\text{where } p = Q\Big(\frac{y[n](1-C_\delta)}{\sigma_1}\Big) \tag{34}$$

*Proof.* First we observe that the small imaginary component $x_2$ of the perturbation mostly rotates the received signal, with most of the increase in $y[n+1]$ coming from $x_1$ (see Fig. 4).

$$\begin{aligned}\mathrm{Mag}(\bar{\phi}+\bar{\delta}) &= \big|C_\delta y[n] + x_1 + jx_2\big| \\ &\approx \big(C_\delta y[n] + x_1\big)\end{aligned} \tag{35}$$

Defining $p$ as the probability that $\mathcal{Y}_{best}[n+1] > y[n]$, (33), (34) readily follow from (35), (26) using Gaussian statistics. □

We can rewrite (33) as:

$$\begin{aligned} y[n+1] &= F\big(y[n]\big) = y[n] + f\big(y[n]\big) \\ \text{where } f(y) &\doteq \sigma_1 g\Big(\frac{y(1-C_\delta)}{\sigma_1}\Big) \\ \text{and } g(x) &\doteq \frac{1}{\sqrt{2\pi}} e^{-\frac{x^2}{2}} - xQ(x) \end{aligned} \tag{36}$$

Proposition 2 does not yet allow us to compute the $y[n]$ because it involves the variance $\sigma_1$ that depends on the phases $\phi_i$ of the individual sensors. In the next section we present a statistical distribution for $\phi_i$ that allows us to calculate aggregate quantities such as $\sigma_1$ without knowledge of the individual $\phi_i$.

### B. Statistical Characterization of Sensor Channels

The statistical model is based on the assumption that each sensor has a channel to the BS of similar quality, and unknown phase. This means that the $a_i$'s are all approximately equal, and that the initial values of the phases $\phi_i$ are distributed independently[1] and uniformly in $[0, 2\pi)$. In particular, we set $a_i = 1$ for all sensors, which gives $G^{opt} = N$. As the algorithm progresses towards convergence, the values of $\phi_i$ are distributed over a smaller and smaller range. In general, we expect that the distribution $f_\phi(\phi_i)$ of $\phi_i[n]$ depends on the number of sensors $N$, the iteration index $n$, and the distribution of the perturbations $f_\delta(\delta_i)$. In the spirit of the statistical model, we consider large $N$, and look for a class of distributions that approximate $f_\phi(\phi_i)$.

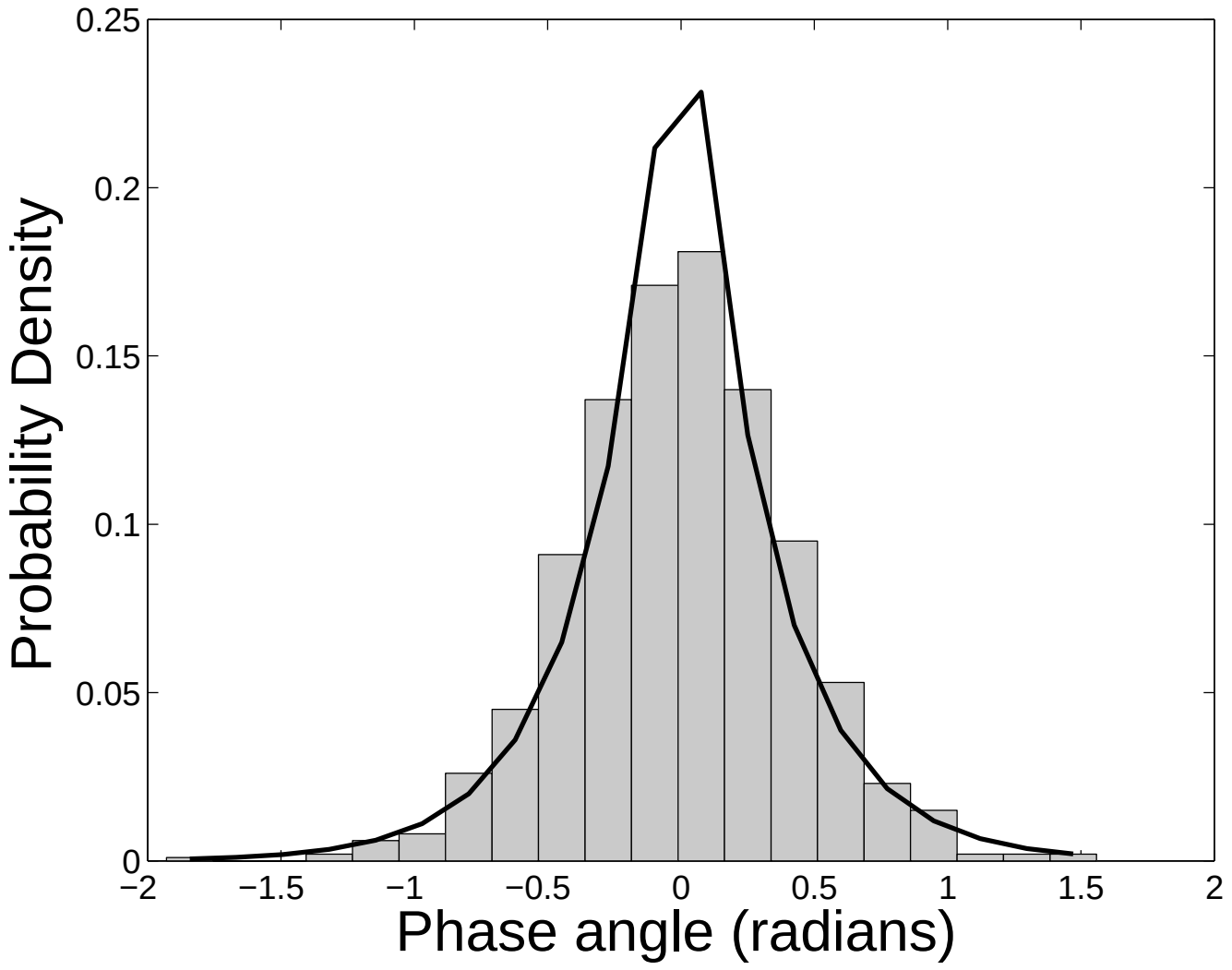

Fig. 5. Comparing a Laplacian Distribution with a Histogram of Empirically Observed Phase Angles

We find that the Laplacian probability distribution gives the best results[2] in terms of accurately predicting the convergence behavior of the algorithm of Section III-A. Fig. 5 shows an empirically derived histogram from a Monte-Carlo simulation of the feedback control algorithm. A Laplacian approximation is also plotted alongside the histogram. We now explain the details of the approximation.

[1] It is important to note that the $\phi_i$ are not *random* variables, however we statistically parametrize them using a probability distribution for the sake of compactness.

[2] The Laplacian distribution for $\phi_i$ is *empirically* found to work well, when compared with other families of distributions like the uniform and triangular distributions.

The Laplacian density function is given by [11]:

$$f_\phi(\phi_i) = \frac{1}{2\phi_0} e^{\frac{-|\phi_i|}{\phi_0}} \tag{37}$$

For $\phi_i$ distributed according to (37), we also have:

$$E\big(\cos\phi_i\big) = \frac{1}{1+\phi_0^2} \tag{38}$$

$$E\big(\cos 2\phi_i\big) = \frac{1}{1+4\phi_0^2} \tag{39}$$

Therefore given that at iteration $n$ of the feedback algorithm, the phase angles are $\bar{\phi}[n] = [\phi_1\phi_2...\phi_N]$, we have:

$$y[n] = \text{Mag}(\bar{\phi}[n]) = \sum_i a_i e^{j\phi_i} \equiv \sum_i \cos\phi_i \tag{40}$$

where we used $a_i = 1$ in (10). Now if we parametrize all the $\phi_i$ using a Laplacian distribution, we can set $\phi_0$ such that $\sum_i \cos\phi_i \equiv NE_\phi\big(\cos\phi\big)$. Thus we use (38) to rewrite (40) as:

$$y[n] = \frac{N}{1+\phi_0^2} \tag{41}$$

We are now able to determine $\sigma_1$ given $y[n]$.

**Proposition 3:** The variance $\sigma_1^2$ of $x_1$ is given by:

$$\sigma_1^2 = \frac{N}{2}\Big((1-C_\delta^2) - \frac{\frac{y[n]}{N}}{4-3\frac{y[n]}{N}}(C_\delta^2 - C_{2\delta})\Big) \tag{42}$$

*Proof.* Equation (42) follows using (32), and the value of the Laplacian parameter from (41) along with the observation that $\sum_i \cos(2\phi_i) = NE_\phi\big(\cos(2\phi)\big) = \frac{N}{1+4\phi_0^2}$. □

Using Propositions 2 and 3, we are able to analytically derive the average convergence behavior of the feedback control algorithm. In particular, we recursively calculate $y[n]$ by substituting the variance $\sigma_1^2$ from (42) into (33).

### C. *Summary of Analytical Model*

We now summarize the analytical model derived in Sections IV-A and IV-B. Our objective is to model the increase over time of the received signal strength by averaging over all possible values of the random perturbations. As mentioned before, we set the channel attenuations for each sensor to unity i.e. $a_i = 1$.

1) Initially we set the received signal strength as $y[1] = \sqrt{N}$. This is the expected value of the signal strength if the initial phase angles are all chosen independently in $[0, 2\pi)$.
2) At each time-interval (iteration) $n > 1$, given the probability distribution of the perturbations $f_\delta(\delta_i)$ and the value of $y[n]$, we compute the Laplacian parameter $\phi_0$ using (41), and then compute the Gaussian variance $\sigma_1^2$ using (42) and finally $y[n+1]$ using the Gaussian statistics in (35) and (26).

## V. Performance Analysis of Feedback Control Protocol

We now present some results obtained from the analytical model of Section IV. Fig. 6 shows the evolution of $y[n]$ derived from the analytical model and also from a Monte-Carlo simulation with $N = 100$, for two different choices of the distribution $f_\delta(\delta_i)$: a uniform distribution in $[-\frac{\pi}{30}, \frac{\pi}{30}]$ and a distribution choosing $\pm\frac{\pi}{30}$ with equal probability. The close match observed between the analytical model and the simulation data provides validation for the analytical model.

We observe from Fig. 6, that the received signal grows rapidly in the beginning, but after $y[n]$ gets to within about 25% of $G^{opt}$, the rate of convergence becomes slower. Also while the simple two-valued probability distribution appears to give good results, it does not satisfy the condition for asymptotic coherence derived in Section III-B.

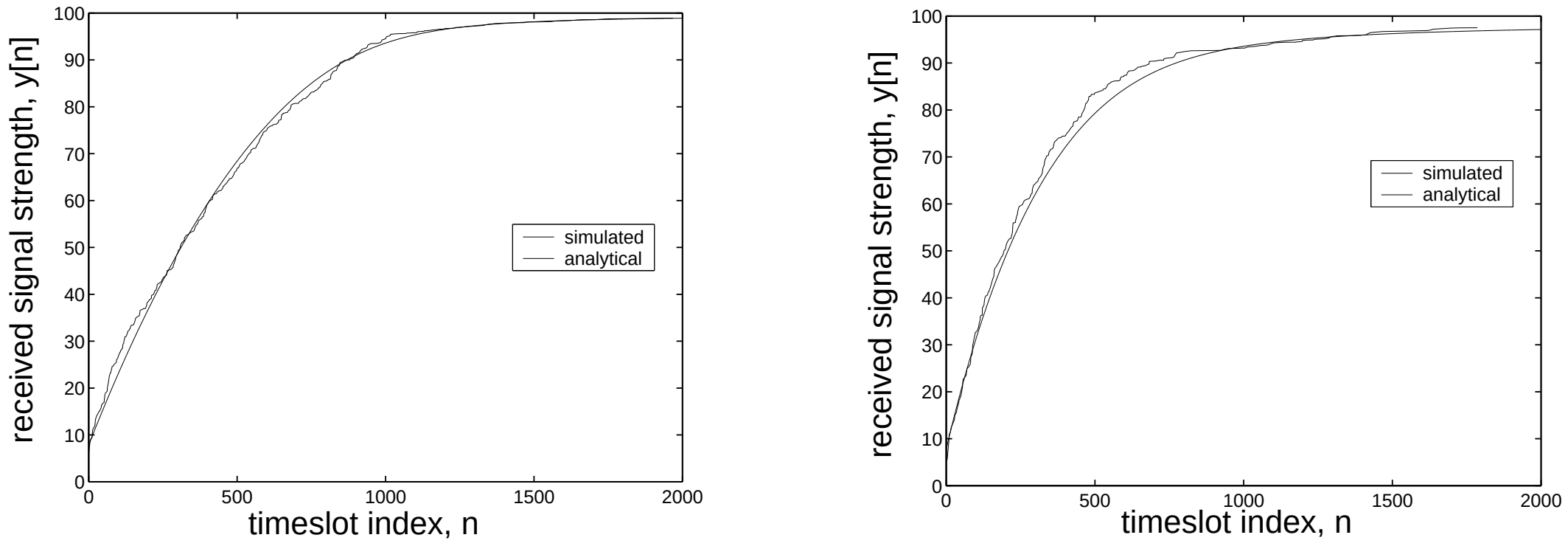

(a) $f_\delta(\delta_i) \sim \mathrm{uniform}(-\frac{\pi}{30}, \frac{\pi}{30})$

(b) $f_\delta(\delta_i) \sim \pm\frac{\pi}{30}$

Fig. 6. Comparison of Analytical Model with Monte-Carlo Simulation of Feedback Control Algorithm

### *A. Optimizing the Random Perturbations*

In Fig. 6, we used the same distribution for the perturbations for all iterations of the algorithm. However this choice is not optimal: intuition suggests that it is best to choose larger perturbations initially to speed up the convergence and make the distribution narrower when the phase angles are closer to coherence. We now use the analytical model to dynamically choose the distribution $f_\delta(\delta_i)$ as a function of $y[n]$. The general problem of choosing a distribution is a problem in calculus of variations. Fortunately, it is possible to restrict ourselves to a family of distributions without losing optimality, because the analytical model only depends on the distribution through the two parameters $C_\delta, C_{2\delta}$. Furthermore the parameters $C_\delta$, $C_{2\delta}$ are highly correlated. To see this recall from (31) and (32) the definitions of $C_\delta$ and $C_{2\delta}$ as the expected values of $\cos(\delta_i)$ and $\cos(2\delta_i)$ respectively. Using the identity $\cos(2\delta) = 2\cos^2(\delta) - 1$ and Jensen's Inequality, we can show that $C_{2\delta}$ is constrained by the value of $C_\delta$:

$$2C_\delta^2 - 1 \le C_{2\delta} \le 2C_\delta - 1 \tag{43}$$

We are interested in $\delta_i$ corresponding to small random perturbations i.e. $\delta_i \ll \frac{\pi}{2}$. For such small values of $\delta_i$, (43) allows only a small range of possible values of $C_{2\delta}$. Indeed we observe that $\cos(\delta_i)$ and $\cos(2\delta_i)$ are very well approximated by the first two terms of the Taylor series:

$$\cos(\delta) \approx 1 - \frac{\delta^2}{2}, \text{ if } |\delta| \ll \frac{\pi}{2} \tag{44}$$

Equation (44) indicates that both $C_\delta$ and $C_{2\delta}$ are essentially determined by the second moment of $\delta_i$, and therefore even a one-parameter family of distributions $f_\delta(\delta_i)$ is sufficient to achieve optimality of the convergence rate. Fig. 7 shows plots of the optimal choices of the $(C_\delta, C_{2\delta})$ pair with $N = 2000$ over $10000$ timeslots for two families of distributions: (i) the 3-point distributions $P(\pm\delta_0) = p, P(0) = 1 - 2p$ parameterised by the pair $(\delta_0, p)$, and (ii) the distributions $\mathrm{uniform}[-\delta_0, \delta_0]$ parametrised by $\delta_0$. At each iteration of the protocol, we used the analytical model to compute the value of the parameters (i.e. the pair $(\delta_0, p)$ in case (i) and $\delta_0$ in case (ii)) that maximizes the $y[n+1]$ given $y[n]$; the optimal parameters in each case were determined numerically using a simple search procedure. The two curves in Fig. 7 were obtained by plotting $(C_\delta, C_{2\delta})$ pair corresponding to the optimal parameters for cases (i) and (ii) at each timeslot. The 3-point distribution is flexible enough to permit any $(C_\delta, C_{2\delta})$ in the feasible region of (43). For the example of Fig. 7, it is clear that the uniform distribution achieves values of $(C_\delta, C_{2\delta})$ that is close to optimal, thereby confirming the intuition of (44).

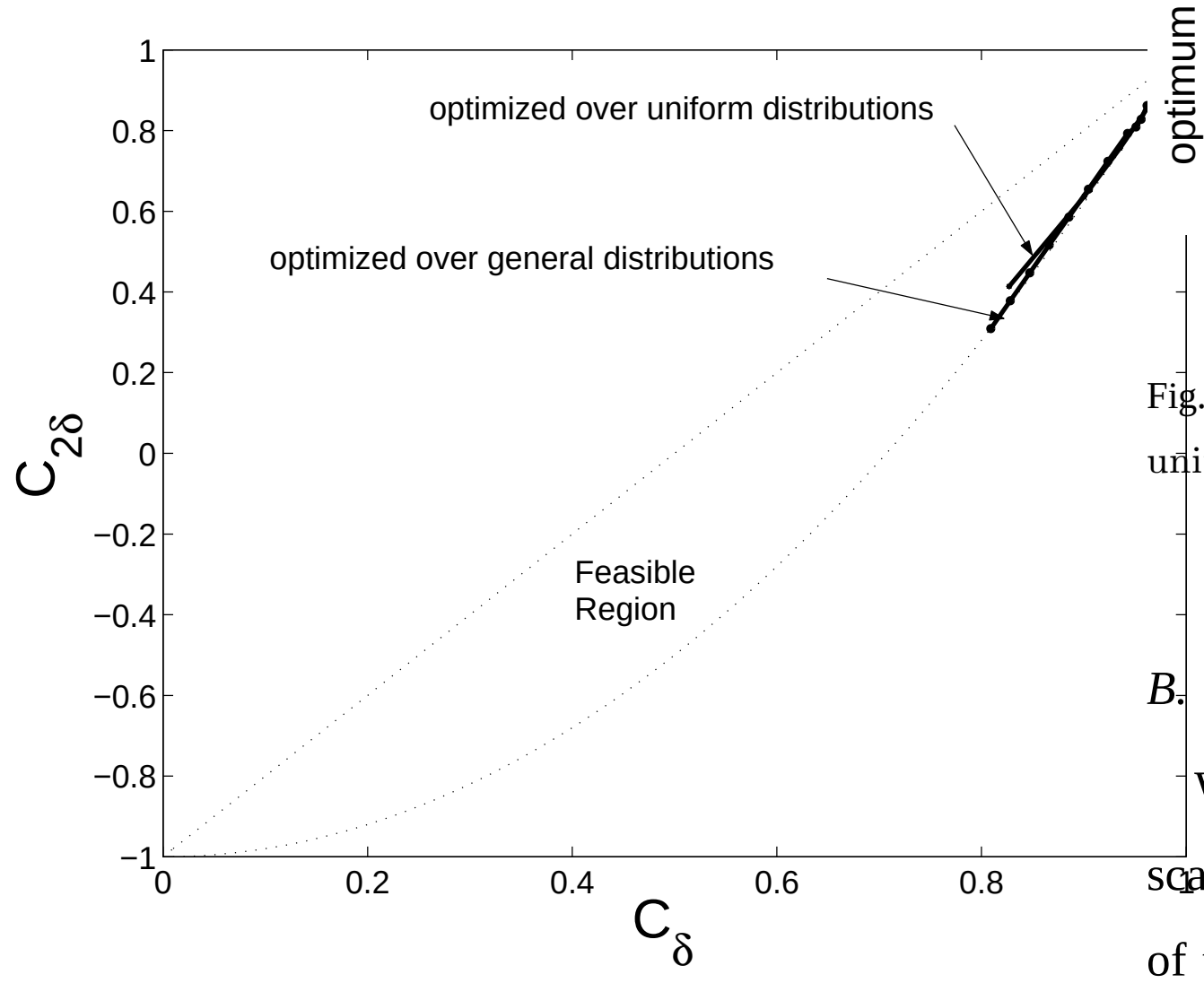

Fig. 7. Near-Optimality of a One-Parameter Distribution

We now use the family of distributions $f_\delta(\delta_i) \sim \text{uniform}[-\delta_0, \delta_0]$ to obtain insight into the optimal convergence rate. Fig. 8 shows $y[n]$ as a function of $n$ for fixed values of $\delta_0$ as well as for the optimized algorithm. We observe that the convergence rate decreases with time in all cases, and the optimized algorithm converges significantly faster than any fixed instance. Fig. 8 also shows the variation of optimal $\delta_0$ with time. This confirms our intuition that at the initial stages of the algorithm, it is preferable to use larger perturbations (corresponding to large $\delta_0$), and when $y[n]$ gets closer to $G^{opt}$, it is optimum to use narrower distributions (corresponding to smaller $\delta_0$).

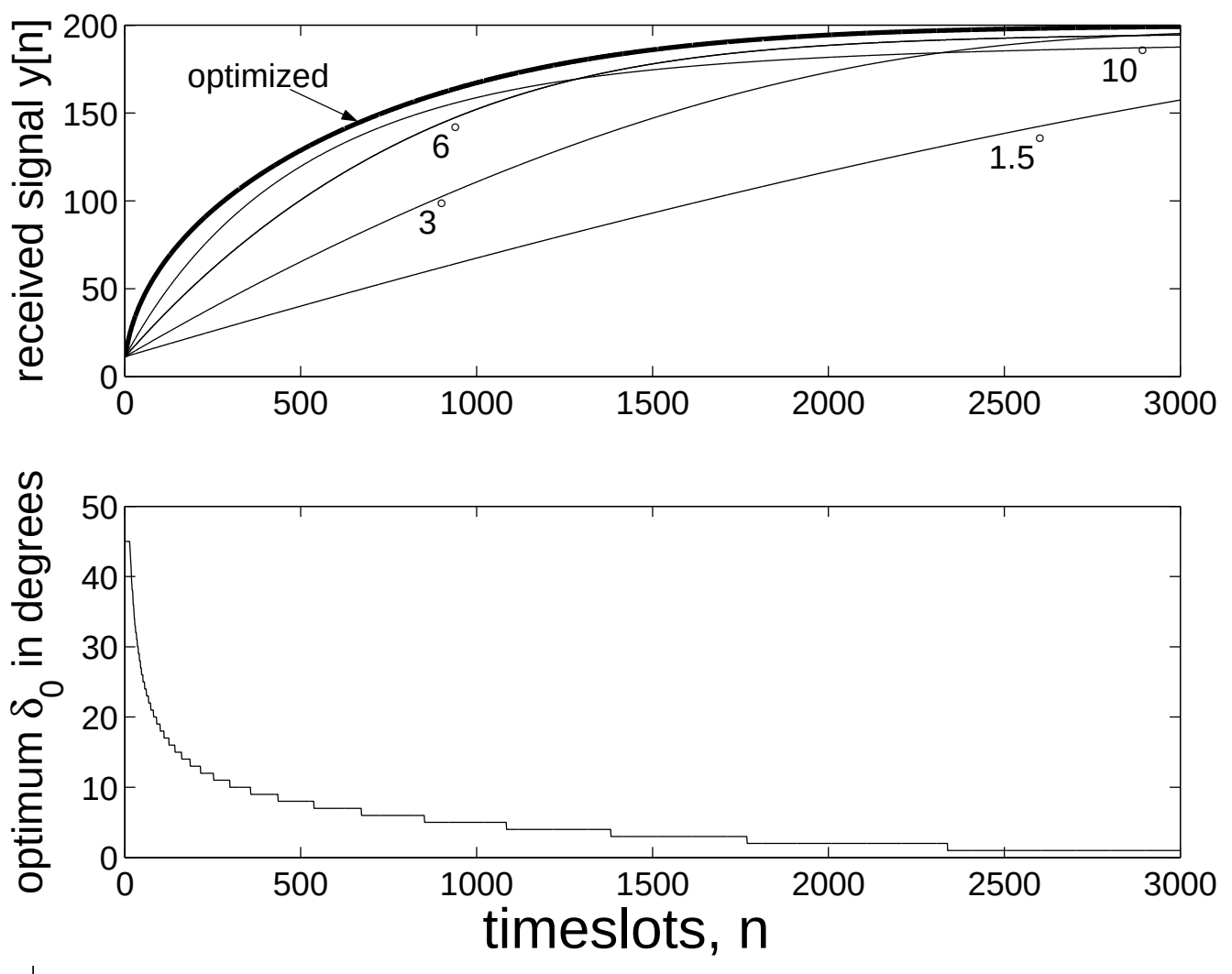

Fig. 8. Optimized algorithm compared to fixed $f_\delta(\delta_i) \sim \text{uniform}[-\delta_0, \delta_0]$ for different $\delta_0$ and $N = 200$

## *B. Scalability Results*

We now turn to the analytical model to study the scalability of the feedback algorithm with the number of transmitting sensors $N$. We show the following scalability results:

- The expected received signal strength at any time, always increases when more transmitters are added.
- The number of timeslots required for the expected signal strength to reach within a certain fraction of convergence always increases with more transmitters, but increases no faster than linearly in the number of transmitters.

**Theorem 2:** Let $y_1[n]$ and $y_2[n]$ be the expected received signal magnitude at timeslot $n$ when the number of transmitting sensors is $N_1$ and $N_2$ respectively. If the sensors use the same distributions $f_\delta(\delta_i)$ for all timeslots $n$, and $N_2 > N_1$, then the following holds for all $n$:

$$y_2[n] \geq y_1[n] \tag{45}$$

$$\text{and } \frac{y_1[n]}{N_1} \geq \frac{y_2[n]}{N_2} \tag{46}$$

*Proof.* We offer a proof by induction. From Section IV-C, we know that $y_2[1] > y_1[1]$ and $\frac{y_1[1]}{N_1} > \frac{y_2[1]}{N_2}$. To prove (45), we need to show that $y_2[n+1] > y_1[n+1]$ given $y_2[n] > y_1[n]$.

We write $y_1[n+1] = F_1(y_1[n]), y_2[n+1] = F_2(y_2[n])$ where $F_1(y)$ and $F_2(y)$ are defined as in (36). Note that $F_1(y_1[n])$ (and $F_2(y_2[n])$) depends on the time index $n$ not only through $y_1[n]$ ($y_2[n]$), but also through the distribution $f_\delta(\delta_i)$. We have suppressed this additional time-dependence to keep the notation simple. The functions $F_1(y)$ and $F_2(y)$ satisfy the following properties:

$$F_2(y) > F_1(y), \forall y \tag{47}$$

$$F_1(y^+) > F_1(y^-), \text{ and } F_2(y^+) > F_2(y^-) \text{ if } y^+ > y^- \tag{48}$$

To see this we observe from (42) that for the same value of $y$, $\sigma_1$ is larger for larger $N$, and since $f(y)$ in (36) increases with $\sigma_1$, (47) follows. To show (48), it is sufficient to show that $F_1(y)$ and $F_2(y)$ have a positive derivative with respect to $y$. This can be shown readily by differentiating the expression in (36):

$$\frac{dF_1(y)}{dy} = \frac{d}{dy}\big(y+f(y)\big) = 1-(1-C_\delta)Q\Big(\frac{y(1-C_\delta)}{\sigma_1}\Big) > C_\delta > 0 \tag{49}$$

We are now ready to complete the proof of (45) by induction. Given that $y_2[n] > y_1[n]$, we have:

$$y_2[n+1] \equiv F_2\big(y_2[n]\big) > F_1\big(y_2[n]\big) > F_1\big(y_1[n]\big) \equiv y_1[n+1] \tag{50}$$

where we used (47) and (48) for the two inequalities.

This completes the proof of (45). The proof of (46) by induction is similar and is omitted. □

**Corollary:** The scalability relations (45) and (46) hold when the sensors use optimized distributions $f_\delta(\delta_i)$ in both cases.

*Proof.* Let $\tilde{y}_1[n]$ and $\tilde{y}_2[n]$ be the expected received signal magnitudes using the respective optimized distributions. We apply Theorem 2 to the case where we use the distribution $f_\delta(\delta_i)$ optimized for $N_1$ sensors in both cases. By definition $\tilde{y}_2[n] \geq y_2[n]$, and $\tilde{y}_1[n] = y_1[n]$, therefore $\tilde{y}_2[n] \geq \tilde{y}_1[n], \forall n$. This proves (45). Using the same argument for the distribution $f_\delta(\delta_i)$ optimized for $N_2$ sensors, we can prove (46). □

Another important criterion for scalability is the number of timeslots $T_f(N)$ required for the algorithm to converge to a fixed fraction, say $f = 0.75$ or 75% of the maximum for $N$ transmitting sensors. Theorem 2 shows that $T_f(N)$ is an increasing function of $N$. Next we show that when the feedback algorithm is appropriately optimized, $T_f(N)$ increases with $N$ no faster than linearly.

**Theorem 3:** The number of timeslots to convergence satisfies the following:

$$\lim_{N\to\infty} \frac{T_f(N)}{N} \leq t_f, \text{ where } t_f \text{ is some constant.} \tag{51}$$

*Proof.* First we use (43) to get a lower-bound for the variance $\sigma_1^2$. With $y[n] = f\cdot G^{opt} = f\cdot N$ we have:

$$(1-C_\delta)^2 \leq C_\delta^2 - C_{2\delta} \leq (1-C_\delta^2) \tag{52}$$

Using the upper bound from (52) in (42), we have:

$$\begin{aligned}\sigma_1^2 &> N\frac{(1-C_\delta^2)}{2}\Big(\frac{4N-4y[n]}{4N-3y[n]}\Big)\\ &> 2N(1-C_\delta)\Big(\frac{1-f}{4-3f}\Big)\end{aligned} \tag{53}$$

We now use a bound for the Gaussian Q-Function:

$$Q(x) > \frac{1}{\sqrt{2\pi}}e^{-\frac{x^2}{2}}\big(\frac{1}{x} - \frac{1}{x^3} + \frac{3}{x^5}\big) \tag{54}$$

Using (54), we rewrite (33) to get:

$$\Delta y[n] \doteq y[n+1] - y[n] > \frac{\sigma_1}{\sqrt{2\pi}}e^{-\frac{x^2}{2}}\big(\frac{1}{x^2} - \frac{3}{x^4}\big) \tag{55}$$

$$\text{where } x = \frac{y[n](1-C_\delta)}{\sigma_1} \tag{56}$$

The bound in (55) has a maximum at $x_0 \approx 3.6$; choosing a $C_\delta$ such that $x$ is close to $x_0$, does not necessarily optimize the RHS in (55), because $\sigma_1$ also depends on $C_\delta$. However such a choice for $C_\delta$ does provide a meaningful lower bound on the optimal $\Delta y^*[n]$.

$$\sigma_1 \geq \frac{2x_0}{f}\Big(\frac{1-f}{4-3f}\Big) \tag{57}$$

$$\Delta y^*[n] > \frac{2}{f}\Big(\frac{1-f}{4-3f}\Big)\Big(\frac{1}{\sqrt{2\pi}}e^{-\frac{x_0^2}{2}}\big(\frac{1}{x_0}-\frac{3}{x_0^3}\big)\Big) \tag{58}$$

where (57) is obtained by backsubstituting (56) into (53). Let us denote the RHS of (58) by $K(f)$.

We observe that the lower bound in (58) only depends on the fraction $f = \frac{y[n]}{N}$. Let $T_{f,\Delta f}(N)$ be the number of timeslots required for the feedback algorithm to increase $y[n]$ from a fraction $f - \Delta f$ to a fraction $f$ of convergence. If $\Delta f$ is small enough, we can use (58) to write:

$$\begin{aligned} \Delta f \cdot N &= y[n] - y[n - T_{f,\Delta f}(N)] \\ &= \sum_{t=1}^{T_{f,\Delta f}(N)} \Delta y[n-t] \\ &\approx \Delta y[n] \cdot T_{f,\Delta f}(N) \\ &> K(f) T_{f,\Delta f}(N) \end{aligned} \tag{59}$$

$$\text{Therefore } T_{f,\Delta f}(N) < \frac{\Delta f \cdot N}{K(f)} \tag{60}$$

Since $T_f$ is just a sum of terms like $T_{f,\Delta f}$, (51) immediately follows. □

Theorem 3 is illustrated by the results in Fig. 9, where the number of timeslots required to get within a certain fraction of convergence is plotted against number of transmitters $N$ for a fixed distribution (Fig. 9(a)) as well as optimized distributions (Fig. 9(b)). These results show that the feedback algorithm is highly scalable with number of transmitters.

*C. Tracking Time-varying channels*

So far we have focused on the simple case of time-invariant wireless channels from each sensor to the BS. In practice, the channel phase response varies because of Doppler effects arising from the motion of the sensors or scattering elements relative to the BS. In the distributed beamforming scenario, Doppler effects also arise because of drifts in carrier frequency between the local oscillators of multiple sensors. Therefore an important performance metric for the feedback control algorithm is its ability to track time-varying channels. Intuitively we expect that the algorithm should track well as long as the time-scale of the channel variations is smaller than the convergence time of the algorithm. In light of the scalability results in Section V-B, the algorithm performs better for smaller $N$ because the corresponding convergence time is smaller.

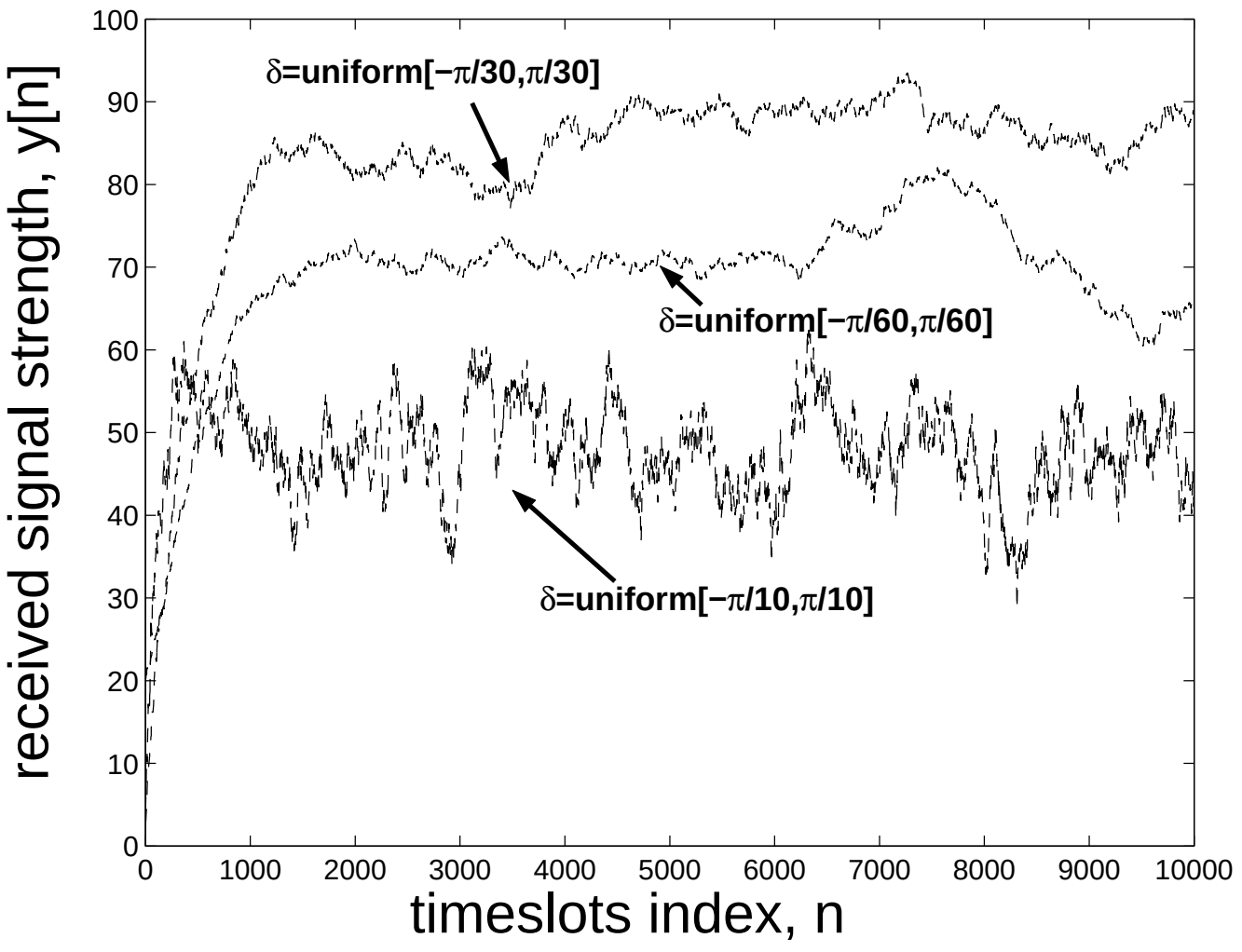

Fig. 10. Tracking Performance for Time Varying Channels: $N = 100$, Doppler rate=$\frac{\pi}{200}$ radians/timeslot

A simulation of $y[n]$ with time in the presence of channel time-variations is shown in Fig. 10. This plot uses a fixed distribution for the phase perturbations, as

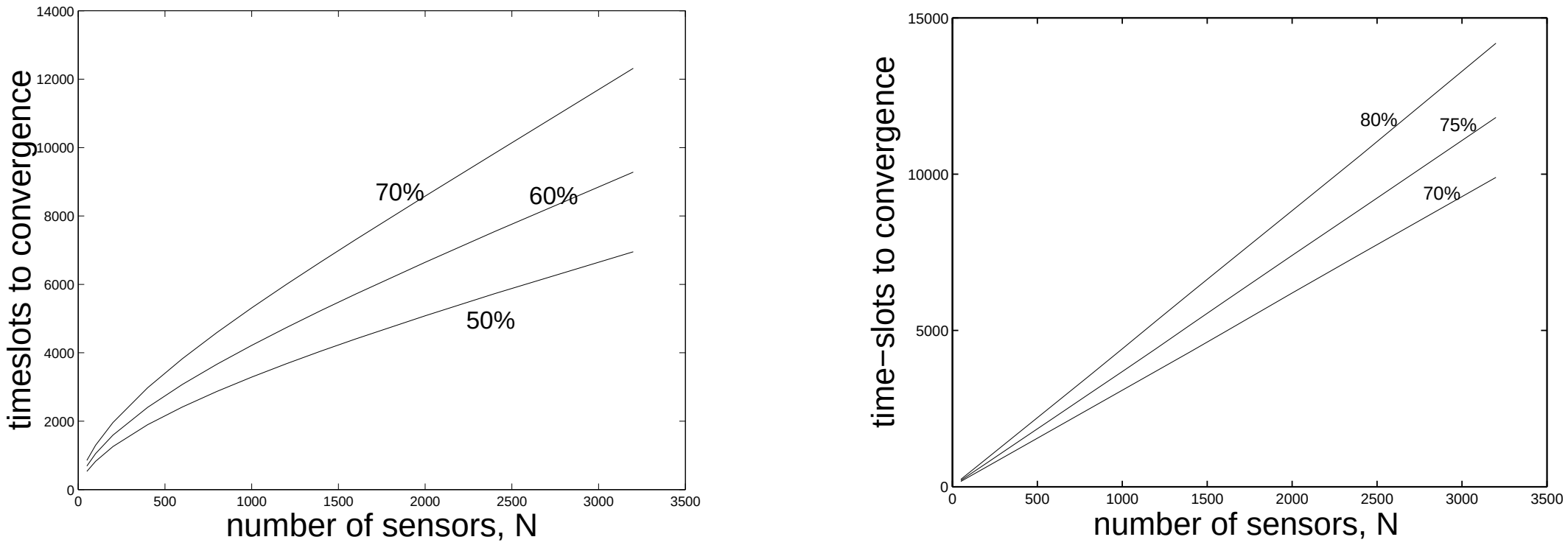

(a) $f_\delta(\delta_i) \sim \mathrm{uniform}(-\frac{\pi}{90}, \frac{\pi}{90})$

(b) Optimized Distribution

Fig. 9. Scalability of Feedback Control Algorithm with Number of Sensors

the analytical model for optimization is not applicable to the time-varying case. A more detailed study of the tracking performance of the feedback control algorithm is beyond the scope of the present work.

## VI. Conclusion

In this paper, we presented a simple algorithm for distributed beamforming in sensor networks, that is based on the idea of using SNR feedback from the receiver to perform phase synchronization in an iterative manner. This algorithm can be easily implemented in a decentralized manner and is guaranteed to achieve asymptotic coherence under mild assumptions. We also derived an analytical model that predicts the performance of the algorithm accurately, and offers insight into the convergence behavior.

This paper represents an initial study into a new approach to the problem of distributed synchronization, and leaves several open issues. We presented the Laplacian distribution to model the statistics of the phase angles $\phi_i$ as an empirical observation. However the underlying reason why the Laplacian distribution works so well is not clear. In addition the stability and convergence behavior of the feedback control algorithm under non-idealities like time-varying channels, and the effects of noise are open issues for future work.

While we use the term “sensors” for the cooperating nodes performing distributed beamforming, the technique developed here is of more general applicability. For example, it could be used as the basis for cooperative communication between clusters of nodes in a wireless ad hoc network. In such a context, it would be of interest to examine how the use of distributed beamforming would impact the design of medium access control and network layer protocols.

## Acknowledgment

The authors would like to acknowledge a discussion with Dr. Babak Hassibi which stimulated detailed investigation of the scalability of the proposed protocol.